\documentclass[12pt]{article}

\newcommand{\beg}{\begin{equation}}
\newcommand{\ene}{\end{equation}}

\begin{document}
\title{
\textsc{\bf Seven Principles of Quantum Mechanics}}

\author{
 Igor V. Volovich
\\ $~~~$\\
\textsf{Steklov Mathematical Institute}\\
\textsf{Russian Academy of Sciences}\\
\textsf{Gubkin St. 8, 117966, GSP-1, Moscow, Russia}\\
\emph{e-mail: volovich@mi.ras.ru} }
\date {~}
\maketitle
\begin{abstract}
 The list of basic axioms of quantum mechanics as it
  was formulated by von
 Neumann  includes only the mathematical formalism of the Hilbert
 space and its statistical interpretation. We point out that such
 an approach is too general to be considered as the foundation of
 quantum mechanics. In particular in this approach any
 finite-dimensional Hilbert space  describes a quantum system.
 Though such a treatment might be a convenient approximation it can
 not be considered as a fundamental description of a quantum system
 and moreover it leads to some paradoxes like Bell's theorem.
  I present a list from seven basic postulates of axiomatic
 quantum mechanics.  In particular the list includes the axiom
 describing spatial properties of quantum system. These axioms do
 not admit a nontrivial realization in the finite-dimensional
 Hilbert space. One suggests that the axiomatic quantum mechanics
is consistent with local realism.
\end{abstract}
\newpage


{\bf INTRODUCTION}\\~\\

Most discussions of  foundations and interpretations of quantum
mechanics take place around the meaning of probability,
measurements, reduction of the state and entanglement. The list of
basic axioms of quantum mechanics as it was formulated by von
Neumann \cite{Neu} includes only  general mathematical formalism
of the Hilbert space and its statistical interpretation, see also
\cite{Seg}-\cite{QT}.
 From this point of view any
mathematical proposition on properties of operators in the Hilbert
space can be considered as a quantum mechanical result. From our
point of view such an approach is too general to be called
foundations of quantum mechanics. We have to introduce more
structures to treat a mathematical scheme as quantum mechanics.

These remarks are important for practical purposes. If we would
agree about the basic axioms of quantum mechanics and if one
proves a proposition in this framework then it could be considered
as a quantum mechanical result. Otherwise it can be a mathematical
result without immediate relevance to quantum theory. An important
example of such a case is related with Bell's inequalities. It is
known that the correlation function of two spins computed in the
four-dimensional Hilbert space does not satisfy the Bell
inequalities. This result is often interpreted as the proof that
quantum mechanics is inconsistent with Bell's inequalities.
However from the previous discussion it should be clear that such
a claim is justified only if we agree to treat the
four-dimensional Hilbert space as describing a physical quantum
mechanical system. In quantum information theory qubit, i.e. the
two-dimensional Hilbert space, is considered as a fundamental
notion.

Let us note however that in fact the finite-dimensional Hilbert
space should be considered only as a convenient approximation for
a quantum mechanical system and if we want to investigate
fundamental properties of quantum mechanics then we have to work
in an infinite-dimensional Hilbert space because only there the
condition of locality in space and time can be formulated. There
are such problems where we can not reduce the infinite-dimensional
Hilbert space to a finite-dimensional subspace.

We shall present a list from seven axioms of quantum mechanics.
The axioms  are well known from various textbooks but normally
they are not combined together. Then, these axioms define an
axiomatic quantum mechanical framework. If some proposition is
proved in this framework then it could be considered as an
assertion in axiomatic quantum mechanics. Of course, the list of
the axioms can be discussed but I feel that if we fix the list it
can help to clarify some problems in the foundations of quantum
mechanics.

For example, as we shall see, the seven axioms do not admit a
nontrivial realization  in the four-dimensional Hilbert space.
This axiomatic framework requires an infinite-dimensional Hilbert
space. One can prove that Bell's inequalities might be consistent
with the correlation function of the localized measurements of
spin computed in the infinite-dimensional Hilbert space \cite{Vol,
Vol2}. Therefore in this sense we can say that axiomatic quantum
mechanics is consistent with Bell's inequalities and with local
realism. It is well known that there are no Bell's type
experiments without loopholes, so there is no contradiction
between Bell's inequalities, axiomatic quantum mechanics and
experiments, see \cite{KhV}.

 There is a gap between an abstract approach to the foundations
and the very successful pragmatic approach to quantum mechanics
which is essentially reduced to the solution of the
Schr$\ddot{o}$dinger equation. If we will be able to fill this gap
then perhaps it will be possible to get a progress in the
investigations of foundations because in fact the study of
solutions of the Schr$\ddot{o}$dinger equation led to the deepest
and greatest achievements of quantum mechanics.

In this note it is proposed that the key notion which can help to
build  a bridge between the abstract formalism of the Hilbert
space and the practically useful formalism of quantum mechanics is
the notion of the ordinary three-dimensional space. It is
suggested that the spatial properties of quantum system should be
included into the list of basic axioms of quantum mechanics
together with the standard notions of the Hilbert space,
observables and states. Similar approach is well known in quantum
field theory but it is not very much used when we consider
foundations of quantum mechanics.

 Quantum mechanics is essentially reduced to the solution of the
Schr$\ddot{o}$dinger equation. However in many  discussions of the
foundations of quantum mechanics not only the Schr$\ddot{o}$dinger
equation is not considered but even the space-time coordinates are
not mentioned (see for example papers in \cite{QT}). Such views to
the foundations of quantum mechanics are similar to the
consideration of foundations of electromagnetism but without
mentioning the Maxwell equations.

 Here I present a list from seven basic
postulates of quantum mechanics which perhaps can serve as a basis
for  further discussions. The axioms are: Hilbert space,
measurements, time, space, composite systems, Bose-Fermi
alternative, internal symmetries. In particular the list includes
the axiom describing spatial properties of quantum system which
play a crucial role in the standard formalism of quantum
mechanics. Formulations of the axioms are based on the material
from \cite{Neu}-\cite{Vol2}.

The main point of the note is this: quantum mechanics is a
physical theory and therefore its foundations  are placed not in
the Hilbert space but  in space and time.

\section{Hilbert space}

To a physical system one assigns a Hilbert space ${\cal H}$. The
observables correspond  to the self-adjoint operators in ${\cal
H}.$ The pure states correspond  to the one-dimensional subspaces
of ${\cal H}.$ An arbitrary state is described by the density
operator, i.e. a positive operator with the unit trace. For the
expectation value $<A>_{\rho}$ of the observable $A$ in the state
described by the density operator $\rho$, we have the Born-von
Neumann formula
$$
<A>_{\rho}=Tr (\rho A)
$$

\section{Measurements}

Measurement is an external intervention which changes the state of
the system. These state changes are described by the concept of
state transformer or instrument. Let $\{\Omega,{\cal F}\}$ be a
measured space where $\Omega$ is a set and ${\cal F}$ is a
$\sigma$-algebra its subsets. A state transformer $\Gamma$ is a
state transformation valued measure $\Gamma =\{\Gamma_B, B\in
{\cal F} \}$ on the measured space. A state transformation
$\Gamma_B$ is a linear, positive, trace-norm contractive map on
the set of trace class operators in ${\cal H}$.

An ideal state transformer $\Gamma =\{\Gamma_i, i=1,2,... \}$
associated with discrete observable $A=\sum_{i=1}^{\infty} a_iE_i$
is given by the Dirac-von Neumann formula
$$
\Gamma_i(\rho)= E_i\rho E_i
$$
if it is known that the measurement outcome is a real number
$a_i.$ Here $E_i$ is the orthogonal projection operator. Similar
formulae hold for the positive operator valued  measure (POVM).

\section{Time}

The dynamics of the density operator $\rho$ and of a state $\psi$
in the Hilbert space which occurs with passage of time is given by
$$
\rho(t)=U(t)\rho U(t)^*,
$$
$$
\psi (t)=U(t)\psi
$$
Here $t$ is a real  parameter (time), $U(t)$ is a unitary operator
satisfying the abstract Schrodinger equation
$$
i\hbar \frac{\partial}{\partial t}U(t)=H(t)U(t)
$$
where $H(t)$ is the (possibly time dependent) self-adjoint energy
operator (Hamiltonian) and $\hbar$ the Planck constant.

\section{Space}

There exists the three-dimensional Euclidean space ${\bf R}^3$.
Its group of motion is formed by the translation group $T^3$ and
the rotation group $O(3)$. One supposes that in the Hilbert space
${\cal H}$ there is a unitary representation $U(a)$ of the
translation by the three-vector $a.$  If $(\Omega,{\cal F})$ is a
measured space and $\{E_B, B\in {\cal F}\}$ is the associated POVM
then one has
$$
U(a)E_BU(a)^*=E_{\alpha_a (B)}
$$
where $\alpha_a :{\cal F}\to {\cal F}$ is the group of
automorphisms.

One has also a projective representation of the rotation group
$SO(3)$ which can be made into a unitary representation $U(R)$ of
the covering group $SU(2)$, here $R\in SU(2).$ Hopefully the
distinction by the type of argument of $U$ will be sufficient to
avoid confusion. The irreducible representations of $SU(2)$
describes systems with integer and half-integer spins.

\section{Composite systems}

If there are two different systems with assigned Hilbert spaces
${\cal H}_1$ and ${\cal H}_2$ then the composite system is
described by the tensor product
$$
{\cal H}={\cal H}_1\otimes{\cal H}_2.
$$

\section{Bose-Fermi alternative}

The Hilbert space of an $N$-particle system is the $N$-fold tensor
product of the single particle Hilbert spaces provided that the
particles are not of the same species. For identical particles
with integer spin (bosons) one uses the symmetrized $N$-fold
tensor product $({\cal H}^{\otimes N})_S$  of the Hilbert space
${\cal H}.$ For identical particles with half integer spin
(fermions) one uses the anti-symmetrized $N$-fold tensor product
$({\cal H}^{\otimes N})_A$.

\section{Internal symmetries}

There is a compact Lie group $G_{int}$ of internal symmetries and
its unitary representation $U(\tau), \tau \in G_{int}$ in the
Hilbert space ${\cal H}$ which commutes with representations of
the translation group $U(a)$ and the rotation group $U(R).$ For
instance one could have the gauge group $G_{int}=U(1)$ which
describes the electric charge.
The group generates the superselection sectors.\\

{\bf SUMMARY}\\

Axiomatic quantum mechanics described by the presented seven
axioms can be briefly formulated as follows. There is space and
time ${\bf R^1}\times {\bf R^3}$, the symmetry group $G=T^1\times
T^3\times SU(2)\times G_{int}$, the Hilbert space ${\cal H}$ and
the unitary representation $U(g)$ of $G$, here $g\in G.$ Axiomatic
quantum mechanics is given by the following data:
$$
\{{\cal H}, U(g), \rho, (\Omega,{\cal F}, \alpha_g),
\{E_B,\Gamma_B, B\in {\cal F}\}\}
$$
Here $\rho$ is the density operator, $(\Omega,{\cal F})$ is the
measured space , $\alpha_g$ is the group of automorphisms of the
$\sigma$-algebra ${\cal F}$, $\{E_B\}$ is POVM, and $\{\Gamma_B\}$
the state transformer.

 {\bf Example.} An example of quantum system
satisfying the all seven axioms is given by the non-relativistic
spin one half particle with the Hilbert space ${\cal H}={\bf
C}^2\otimes L^2({\bf R}^3)$ and the Schr$\ddot{0}$dinger-Pauli
Hamiltonian and also by
its multi-particle generalization.\\

{\bf COMMENTS}\\~\\
We can add more axioms, of course. In particular we did not
postulate yet the covariance under the Poincare or  Galilei group
(for the Galilei group one has a projective representation) but
only invariance under spatial translations and rotations which we
have in the non-relativistic theory as well as in the relativistic
theory. We could add also the condition of the positivity of
energy. Finally, we could postulate the standard non-relativistic
Schr$\ddot{o}$dinger equation for $N$ bodies  as a fundamental
axiom of quantum mechanics. Note also that in relativistic quantum
field theory all the axioms are valid (in fact we have to add more
axioms to get quantum field theory) except the second axiom on
measurements which requires a special discussion.

Note in the conclusion that the spatial approach to quantum
mechanics explicitly formulated in this note was used for the
investigation of quantum non-locality. It was shown in
\cite{Vol},\cite{Vol2} that quantum non-locality in the sense of
Bell there exists only because the spatial properties of quantum
system were neglected. If we take the spatial degrees of freedom
into account then local realism might be consistent with quantum
mechanics and with performed experiments. If somebody wants to
depart from local realism then he/she has to change quantum
mechanics. The local realist representation in quantum mechanics
was formulated in \cite{Vol2} as follows:
$$
Tr(\rho A_m(x)B_n(y))=E\xi_m(x)\eta_n(y)
$$
Here $A$ and $B$ are observables depending on the space points $x$
and $y$ and on  parameters $m$ and $n$ while $E$ is the
expectation of two random fields $\xi_m(x)$ and $\eta_n(y)$. The
representation was proved in \cite{Vol2} under some rather
restrictive assumptions. It would be important to prove the
representation under more general assumptions. The non-commutative
spectral theory related with the local realist representation is
discussed in \cite{Vol2}, definitions of   local realism in the
sense of Bell and in the sense of Einstein and its relations with
the contextual approach are considered in \cite{KhV}.
\\~\\
{ \bf ACKNOWLEDGMENTS}
\\~\\
I am grateful to Andrei Khrennikov and Richard Gill for numerous
discussions on axiomatics of quantum mechanics and critical
remarks on the manuscript.
 The paper was
supported by the grant of The Swedish Royal Academy of Sciences on
collaboration with scientists of former Soviet Union, EU-network
on quantum probability and applications and RFFI-0201-01084,
INTAS-990590.

\end{document}